\documentclass[fleqn,12pt,twoside]{article}
\usepackage[headings]{espcrc1}

\usepackage{graphicx}
\usepackage[figuresright]{rotating}

\newcommand{\AmS}{{\protect\the\textfont2
  A\kern-.1667em\lower.5ex\hbox{M}\kern-.125emS}}

 \newcommand\la{\langle}
 \newcommand\ra{\rangle}
 \newcommand\beq{\begin{equation}}
 
 \newcommand\eeq{\end{equation}}
 \newcommand\beqn{\begin{eqnarray}}
 \newcommand\eeqn{\end{eqnarray}}

\def\fm{\,\mbox{fm}}
\def\GeV{\,\mbox{GeV}}

\def\lsim{\mathrel{\rlap{\lower4pt\hbox{\hskip1pt$\sim$}}
    \raise1pt\hbox{$<$}}}         
\def\gsim{\mathrel{\rlap{\lower4pt\hbox{\hskip1pt$\sim$}}
    \raise1pt\hbox{$>$}}}         

\title{Time Evolution of Jets and
Perturbative Color Neutralization}

\author{B.Z.~Kopeliovich\address[vina]{Universidad
Tecnica Federico Santa Maria, Valparaiso, Chile}\address[hd]{Institut
f\"ur Theoretische Physik der Universit\"at,
Heidelberg, Germany},
J.~Nemchik\address[kosice]{Institute of Experimental Physics SAV, Kosice,
Slovakia},
Ivan~Schmidt\addressmark[vina]}

\begin{document}

\maketitle

\begin{abstract}

 In-medium production of leading hadrons in hard reactions, carrying the
main fraction of the jet momentum, involves two stages: (i) the
parton originated from the hard process propagates through the
medium radiating gluons due to the initial hard collision, as well
as to multiple interactions in the medium; (ii) perturbative color
neutralization, e.g. picking up an anti-colored parton produced
perturbatively, followed by evolution and attenuation of the
(pre)hadron in the medium. The color neutralization (or production)
length for leading hadrons is controlled by coherence, energy
conservation and Sudakov suppression. The $p_T$-broadening is a
sensitive and model independent probe for the production length. The
color neutralization time is expected to shrink with rising hard
scale. In particular, we found a very fast energy dissipation by a
highly virtual parton: half of the jet energy is radiated during the
first Fermi. Energy conservation makes the production of leading
hadrons at longer times difficult.

\end{abstract}

\section{Introduction}

Modification of jets produced in a medium is currently a hot topic.
Data for heavy ion collisions at RHIC \cite{star,phenix} show that
hadrons produced with high $p_T$ in central gold-gold collisions at
$\sqrt{s}=200\GeV$ are five times suppressed compared to pp
collisions. This observation is an evidence for a dense medium
created in heavy ion collisions. However, to be confident of the
medium properties resulting from the data analysis one has to employ
a reliable theoretical description of jet development in the medium.
Unfortunately, this is not the case so far. The current theoretical
models relating the observed jet quenching to medium-induced energy
loss are based on unjustified ad hoc assumptions: (i) the high-$p_T$
parton initiating the jet always keeps hadronizing and radiating
gluons outwards the medium; (ii) the effect of the medium can be
reproduced by a simple shift in the argument of the fragmentation
function. In this paper both assumptions are challenged.

It is clear that no reliable information can be extracted from data,
if the analysis is based on a theoretical model which has never been
tested. This hardly can be done in heavy ion collisions in view of
too many uncertainties. The density and space-time development of
the medium are unknown and model dependent, and the jet energy is
also unknown.

\begin{figure}[htb]
\begin{minipage}[t]{60mm}
 \includegraphics[width=59mm]{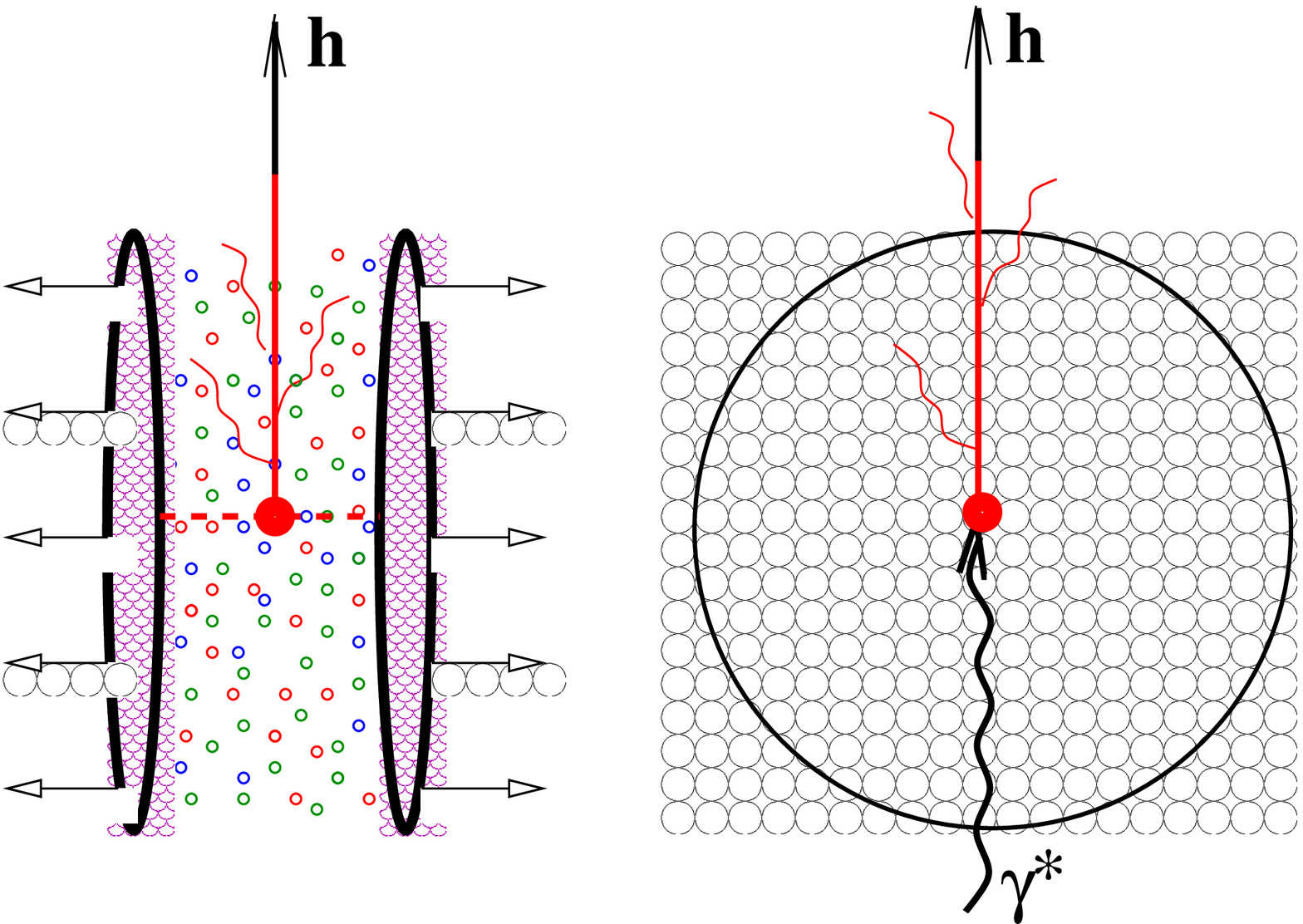}
 \caption{{\it Left:} high $p_T$ hadron production in a heavy ion
collision. The parton propagates through the created medium which modifies
its hadronization. {\it Right:} leading hadron production in DIS on a
nucleus. The nuclear density and the kinematics of the reaction are under
control.}
 \label{pt-dis}
 \end{minipage}
 \hspace{\fill}
\begin{minipage}[t]{85mm}
 \includegraphics[width=84mm]{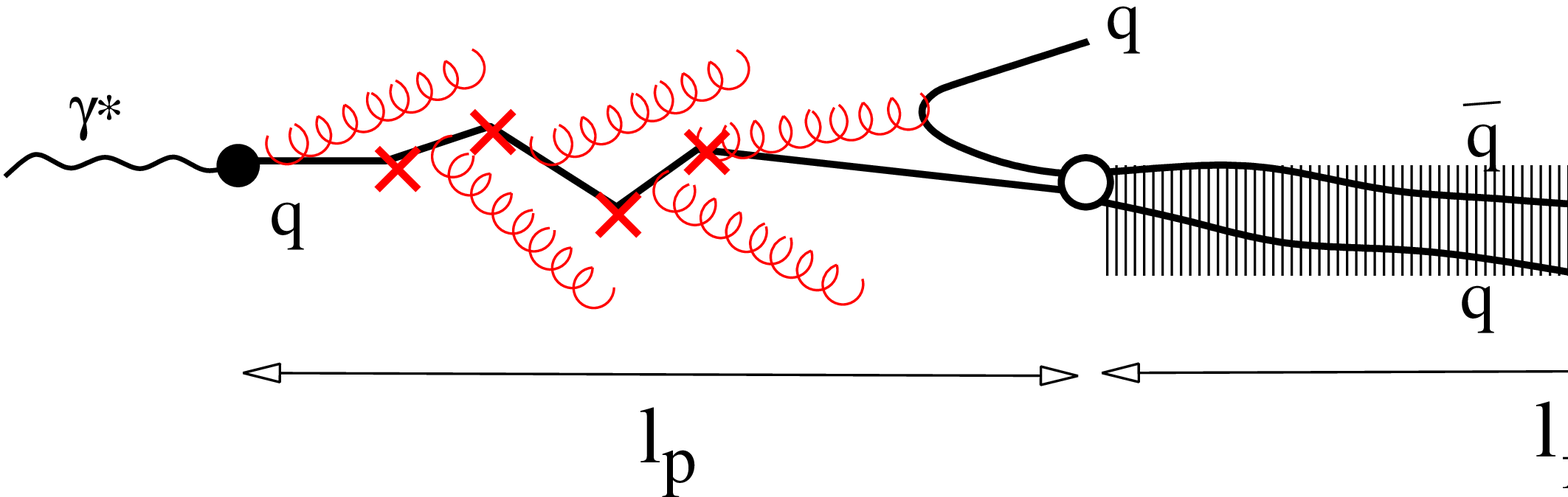}
 \caption{Two-step process of leading hadron production. On the production
length $l_p$ the quark is hadronizing experiencing multiple interactions
broadening its transverse momentum and inducing an extra energy loss.
Eventually the quark color is neutralized by picking up an antiquark. The
produced color dipole (pre-hadron) is attenuating in the medium and
developing the hadron wave function over the formation path length $l_f$.}
 \label{2-step}
 \end{minipage}
 \end{figure}

A process which looks similar, but which has much less
uncertainties, is hadron production in DIS on nuclei, as is
illustrated in Fig.~\ref{pt-dis}.

 In this case the density and geometry of the
medium, the kinematics are known. Moreover, the parton virtuality $Q^2$
and its energy $\nu$ are not correlated allowing diverse tests.

\section{Jet quenching in DIS}

In this case the fraction $z_h=E_h/E_q$ of the jet energy taken by the
hadron is measured, and we are interested in leading hadron production
$z_h\gsim0.5$. The space-time development of the hadronization process
which ends up with production of the leading hadron should have a two-step
structure as is illustrated in Fig.~\ref{2-step} \cite{knp}.

 Since the produced pre-hadron strongly (exponentially) attenuates in the
nuclear medium, the position of the color neutralization point is crucial
for the resulting nuclear suppression. In the energy loss scenario the
color neutralization is assumed to happen always outside the nucleus,
$l_p> R_A$, so that the pre-hadron has no chance to interact.

However, if a large fraction, $z_h\to1$, of the initial quark energy is
taken by the produced hadron, the process of color neutralization cannot
last long because of energy conservation. Indeed, while the quark is
propagating and radiating gluons either in vacuum, or in a medium, it
keeps losing energy and once the quark energy comes below than $z_h E_q$,
there is no chance any more to produce a hadron with fractional energy
$z_h$. Thus, energy conservation imposes a restriction on the color
neutralization time \cite{kn},
 \beq
l_p\leq \frac{E_q}{\la dE/dz\ra}\,(1-z_h)\ ,
\label{10}
 \eeq
 which must vanish at $z_h\to1$.

 All the consideration hereafter is held in the rest frame of the target.
We use here the mean rate of energy loss per unit of length, $z$,
although its time dependence and fluctuations may be important and
should be taken into account \cite{knph}. In the simplest case of
the string model the rate of energy loss is just the string tension,
$-dE/dz\approx 1GeV/fm$ \cite{kn}. In a hard reaction the rate can
be much higher, since a highly virtual parton radiates gluons
intensively. Integrating over the radiation spectrum one arrives at
a time independent rate \cite{feri} which rises quadratically with
the hard scale \cite{knp,knph},
 \beq
-\frac{dE}{dz}= \frac{2\alpha_s}{3\pi}\,Q^2\ .
\label{20}
 \eeq
 Actually, at large $z_h$ one should also impose energy conservation
restrictions on these calculations, as well as introduce a Sudakov factor
which emerges due to this restriction. As a result, Eq.~(\ref{20}) holds
only over the distance $\Delta z<z_1=2(1-z_h)E_q/Q^2$ after the DIS, and
then slows down \cite{knp,knph}.

Thus, the color neutralization time is controlled by the jet energy,
virtuality and fractional energy of the detected hadron as,
 \beq
\la l_p\ra \propto \frac{E_q}{Q^2}\,(1-z_h)\ .
\label{30}
 \eeq

\section{Perturbative color neutralization}

It is widely believed that hadronization and color neutralization
are soft processes with a long duration time controlled by the soft
scale. This is correct, as far as it concerns the formation of the
final hadron wave function,
 \beq
l_f\sim \frac{E_h}{\Lambda^2_{QCD}}\ .
\label{40}
 \eeq
 This seems to be in contradiction with the above estimates for color
neutralization time limited by energy conservation, but this is just
the consequence caused by mixing up two different time scale. The
former, the production time or length, $l_p$, is the distance
covered by the hadronizing parton until its color is neutralized by
another (anti)parton. The colorless system produced is not a hadron
yet, since it has no wave function. For instance, it might be a
$\bar qq$ dipole with a definite separation, which even has no
definite mass and can be projected to wave functions of different
hadrons.  It still needs a rather long time called formation time,
Eq.~(\ref{40}), to form the hadron.

\begin{figure}
\begin{center}
 \includegraphics[width=10cm]{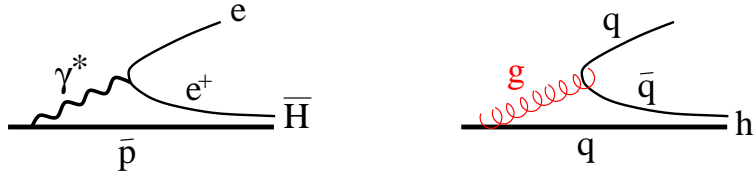}
 \end{center}
 {\caption[Delta] {Perturbative charge neutralization and production of an
antihydrogen in QED \cite{bms} (left figure) versus perturbative color
neutralization and production of a (pre)hadron in QCD
\cite{berger,knp,knph} (right figure).}
 \label{Hbar}}
 \end{figure}

The early color neutralization suggested by Eq.~(\ref{30}) is a
perturbative process. It has been also known in QED, for instance,
in the production of antihydrogen found at CERN, which was predicted
in \cite{bms} as a perturbative neutralization of the antiproton
charge, as is illustrated in Fig.~\ref{Hbar}.
 Color neutralization in QCD looks alike \cite{berger,knp,knph} and leads
to the creation of a pre-hadron, as is shown in Fig.~\ref{Hbar}.

Early perturbative color neutralization is an essential ingredient
of the phenomenon known as color transparency. Production of neutral
pions in $\pi^-A$ collisions at $40\GeV$ was measured in the limit
$z_h\to1$ in \cite{ihep}. A strong signal of color transparency was
detected in agreement with prediction \cite{kz}. This is a direct
proof of the perturbative nature of the produced pre-hadron.

\section{Mapping the color neutralization length with nuclear broadening}

A quark originated from DIS on a free proton may have a transverse
momentum relative the initial photon direction. This happens due to
the intrinsic motion of quarks in the proton. In the case of a
nuclear environment the quark propagating through the medium
increases its transverse momentum due to multiple interactions. This
leads in $p_T$ broadening,
 \beq
\Delta p_T^2 = \la p_T^2\ra_A - \la p_T^2\ra_p\ .
\label{45}
 \eeq
 Notice that an inclusive hadron with transverse momentum $\vec k_T$ is
experimentally observed, rather than the quark. If the hadron
carries fraction $z_h$ of the quark longitudinal momentum, they are
related as $p_T=z_h\,k_T$, when the transverse momentum of the
hadron relative to the quark direction is neglected. This relation,
called sometimes {\it seagull effect}, is well confirmed by data.
Thus, the quark $p_T$ broadening is related to the observed hadronic
broadening as,
 \beq
\Delta p_T^2 = {1\over z_h^2}\,\Delta k_T^2\ .
\label{47}
 \eeq
 This relation does not need the above assumption that the hadron and
quark fly in the same direction. Indeed, the mean transverse
momentum squared between the quark and hadron cancels in (\ref{45})
if the jet shape is not disturbed by the medium. The jet shape might
be affected by the medium, but this is a small, second order
correction.

 From the relation Eq.~(\ref{47}) we conclude that binning of data in
$z_h$ is very important, while data averaged over $z_h$ is difficult to
compare with theory.

After the quark color is neutralized and a colorless pre-hadron is created,
we should forbid further inelastic (color-exchange)  interactions.  One may
think that if the pre-hadron experienced an inelastic collision, the
detected hadron could be recreated again. However, its momentum would be
substantially reduced, i.e. it would feed down the small $z_h$ part of the
spectrum. Direct calculations confirm that the contribution of such a
two-fold interaction is indeed vanishingly small at large $z_h$.

Elastic interactions of the pre-hadron are possible, but hardly happen.
Indeed, even for pion the mean free path for elastic collisions is longer
than $20\fm$, and it is $\la r_\pi^2\ra^2/r^4$ times longer for a
pre-hadron of transverse size $r$.

We conclude that the main source of transverse momentum broadening is the
multiple interaction of the original quark during the hadronization stage
along the pathlength $l_p$ (see Fig.~\ref{2-step}. Since broadening is
proportional to the pathlength, it is probably the most direct way to measure
$l_p$.

Thus, we can check our ideas regarding energy, scale and $z_h$
dependence of $l_p$ using broadening, which according to
Eq.~(\ref{30}) is expected to display the following features:
 \begin{itemize}
 \item $\Delta p_T^2$ must rise proportionally to energy $\nu$, unless
$l_p$ exceeds the nuclear size, then this dependence will saturate
and level off;
 \item $\Delta p_T^2$ is expected to fall steadily down towards
 $z_h=1$,
where it should be zero;
 \item $\Delta p_T^2$ should decrease with rising scale $Q^2$. This is
important to test in DIS, since jets produced at mid rapidities in
hadronic (and heavy ion) collisions have maximal virtuality relative
to their energy, therefore $l_p$ should be especially short
\cite{knph}.
 \end{itemize}
 Theoretical tools for calculation of broadening are currently well
developed and have a high predictive power \cite{hans,dhk,jkt}. The
light-cone dipole approach was able explain in a parameter-free way data on
Cronin effect and correctly predict this effect for RHIC \cite{knst}. In
this approach broadening on a pathlength $L$ is given by a simple relation,
 \beq
\Delta\la p_T^2(L)\ra = 2C(s)\int\limits_0^L dz\,\rho_A(z),
\label{50}
 \eeq
 where $\rho_A(z)$ is the nuclear density as function of longitudinal
coordinate $z$ (and implicitly of impact parameter). The coefficient $C$
reads \cite{dhk,jkt},
 \beq
C(s)=\left.\frac{d\sigma_{\bar qq}(r_T,s)}
{dr_T^2}\right|_{r_T=0}\ .
\label{60}
 \eeq
 The cross section $\sigma_{\bar qq}(r_T,s)$ for a $\bar qq$ dipole
 with separation $r_T$ interacting with a proton, introduced in
 \cite{zkl},
cannot be reliably calculated but it has to be fitted to data. There
are currently several popular parametrization fitted to DIS data for
$F^p_2(x,Q^2)$ and photoabsorption data.

Notice that broadening in Drell-Yan reaction, calculated in
\cite{jkt} and relying on Eq.~(\ref{50}), overestimates data from
the E772 experiment \cite{joel} by a factor of two. However, the
recent revision of the data analysis \cite{pt-new}, including the
brand-new data from the E866 experiment concluded that the
theoretical prediction was correct.

Thus, the measurement of broadening in semi-inclusive hadron
production in DIS on nuclei should provide clear and direct
information on the production length $l_p$. This would be a much
better source of information than nuclear attenuation measured so
far, where different effects may be easily mixed up.

\begin{figure}[htb]
\begin{minipage}[t]{75mm}
 \includegraphics[width=68mm]{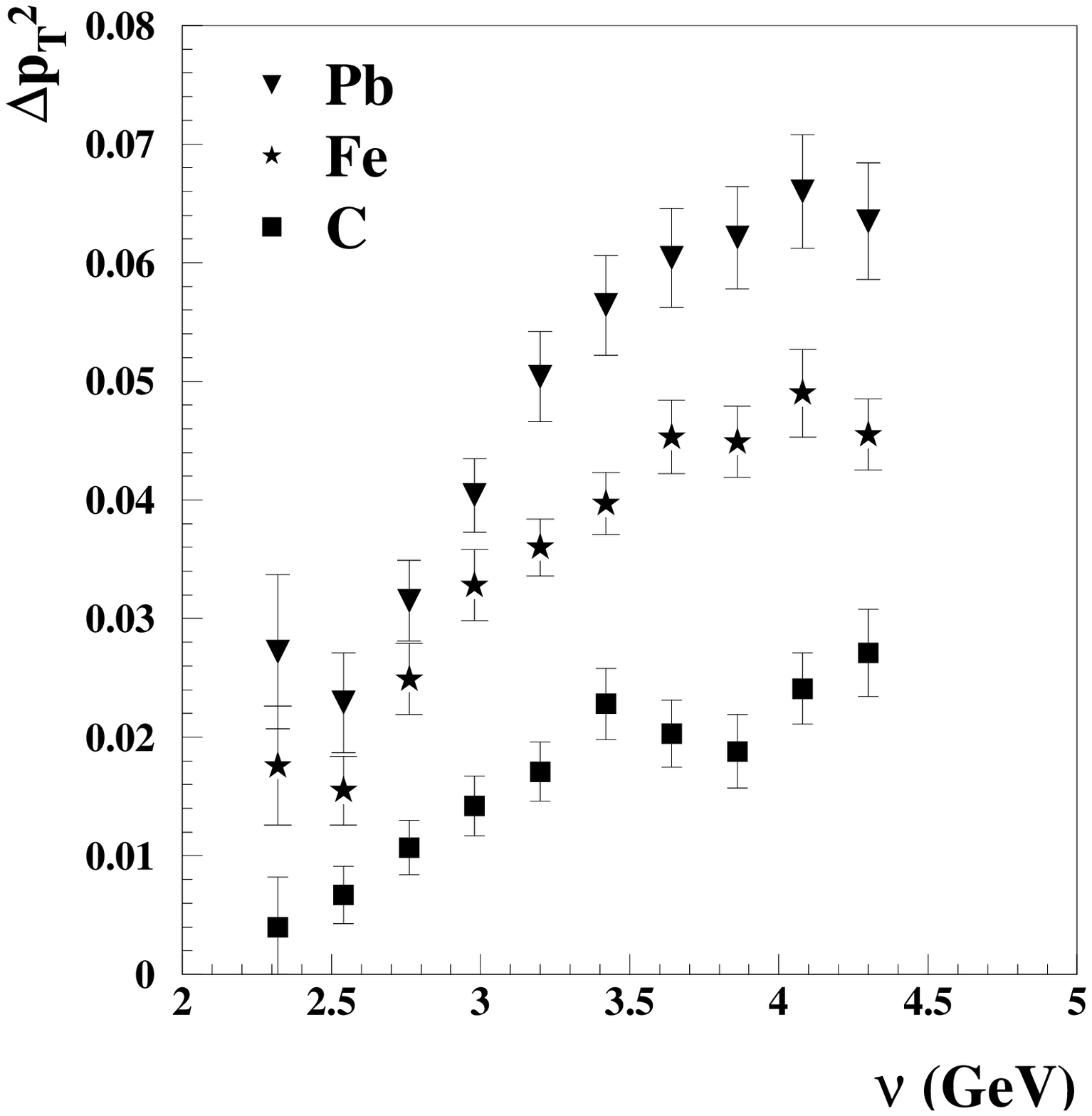}
\caption{Preliminary data from the CLAS EG2 experiment at Jlab \cite{will}
for broadening (in $\GeV^2$) of $\pi^+$ produced with $z_h=0.5-0.6$ in DIS 
on nuclei (Pb, Fe, C from top to bottom) at $Q^2=1-2\GeV^2$ as function of
photon energy $\nu$.}
\label{clas}
\end{minipage}
\hspace{\fill}
\begin{minipage}[t]{7cm}
 \includegraphics[width=65mm]{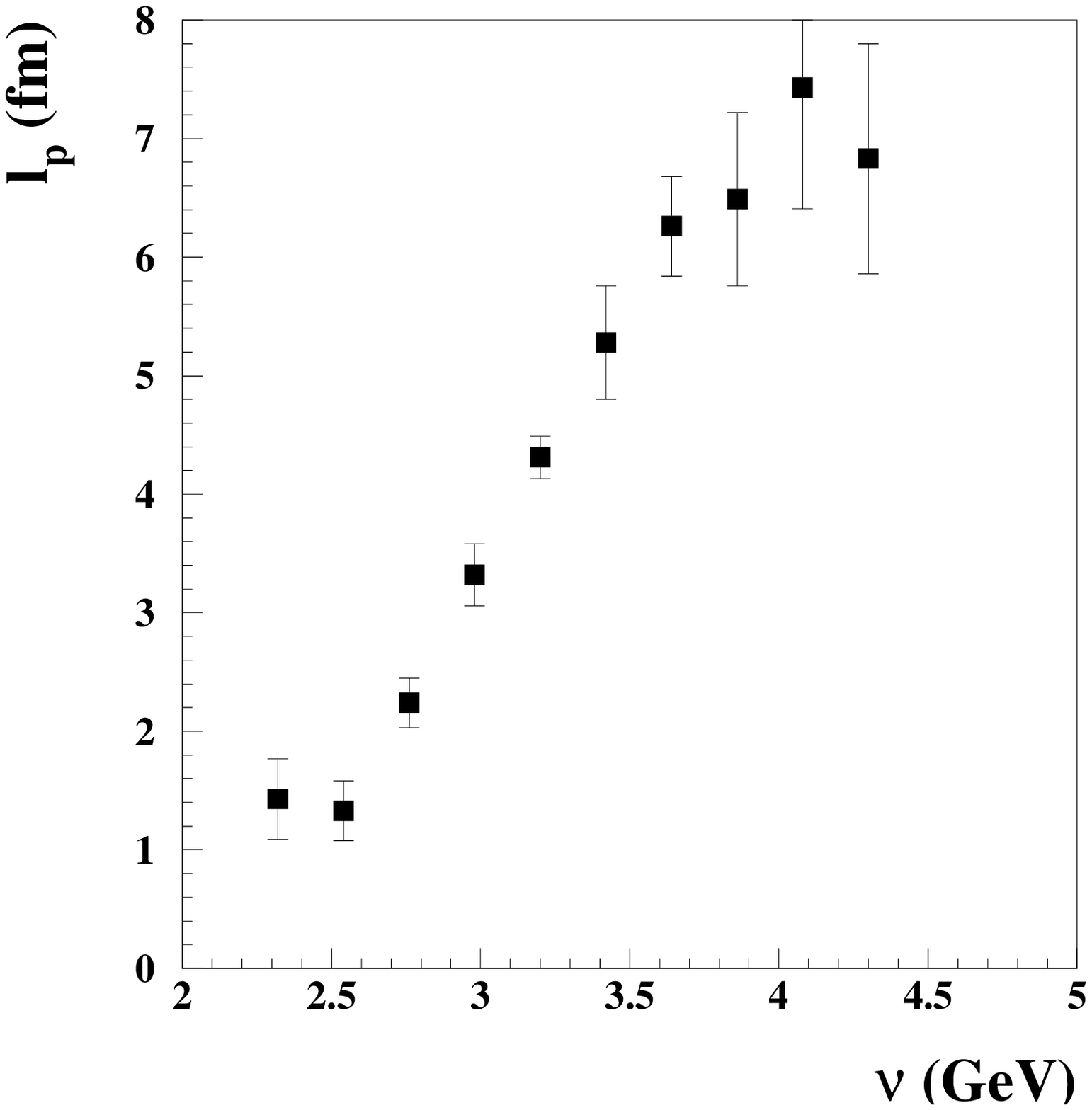}
 \caption{Results of model-independent extraction of the production length
$l_p$ using Eq.~\ref{65} from data depicted in Fig.~\ref{clas}.}
 \label{lp}
\end{minipage}
 \end{figure}

The preliminary data from the CLAS spectrometer at JLab \cite{will}
depicted in Fig.~\ref{clas} indeed demonstrate a steep energy dependence
of broadening in a few GeV region. This confirms the rise of $l_p$ with
energy in accordance with Eq.~(\ref{30}). It also shows that in this range
of energy and $z_h$ the color is neutralized mainly inside the nucleus,
i.e. $l_p<R_A$, otherwise $\Delta p_T^2$ would be energy independent.

Notice that maximal broadening for a quark is about the same as was found 
from Drell-Yan data \cite{pt-new} at much higher energy of $800\GeV$.
This is a surprise, since one should expect factor $C(s)$ to rise with 
energy.

With this data for broadening one can try to extract $l_p$ from data in a
least model-dependent way, using relation,
 \beq
\Delta p_T^2={2Cz_h^2\over A}\int d^2b
\int\limits_{-\infty}^\infty dz\,
\rho_A(b,z)\int\limits_z^{z+l_p}
dz^\prime\,\rho_A(b,z^\prime)
\label{65}
 \eeq
 Comparing this calculation with data one can determine $l_p$ corresponding
to each data point for $\Delta p_T^2$.  Simultaneously, minimizing the
difference between values of $l_p$ on different nuclei one can also
determine factors C, which is not reliably known at such low energies. We
found $C=4.48\pm 0.13$, which is in a good accord with extrapolation from
high energies (see details in \cite{knps}). The results of the analysis
\cite{knps} employing the data for all three nuclei are shown in
Fig.~\ref{lp}.
 This results of model-independent extraction of $l_p$ are to be confronted
with contemporary models for hadronization, those which are based on
perturbative QCD \cite{knp,knph}, or those which employ the ideas of the
string model \cite{kn,acardi}.

It worth reminding that this is the color neutralization length at
$z_h=0.5-0.6$, and one expects a contraction of $l_p$ with rising
$z_h$. Unfortunately no data testing this expectation has been
released so far.

\section{Long production length: upper bound for nuclear modifications}

\begin{figure}[tbh]
\begin{minipage}[t]{70mm}
 \includegraphics[width=60mm]{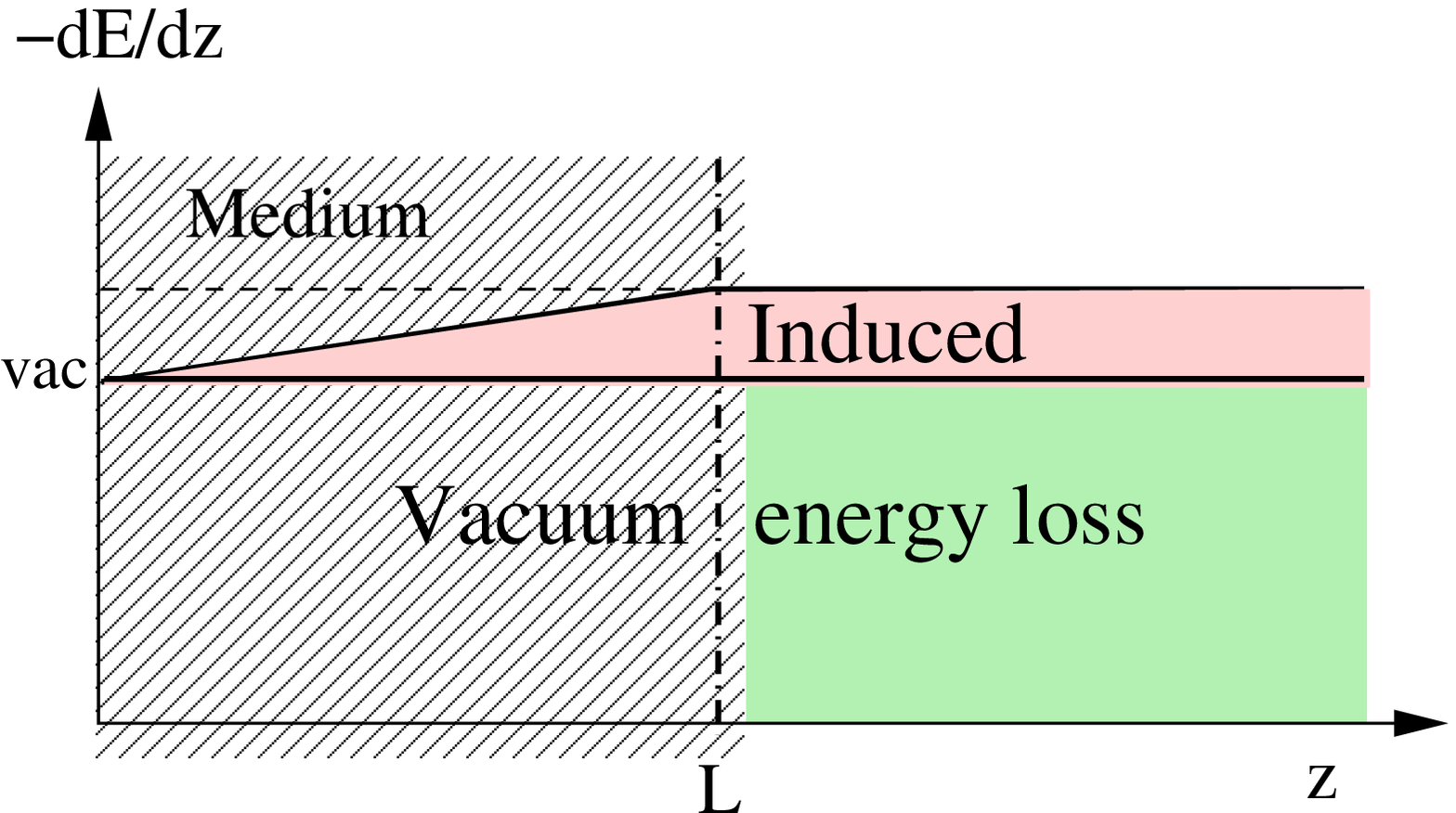}
 \caption{The rate of energy loss following DIS. The induced part rises
linearly with pathlength and then remains constant outside of the medium.}
 \label{eloss1}
\end{minipage}
\hspace{\fill}
\begin{minipage}[t]{70mm}
 \includegraphics[width=60mm]{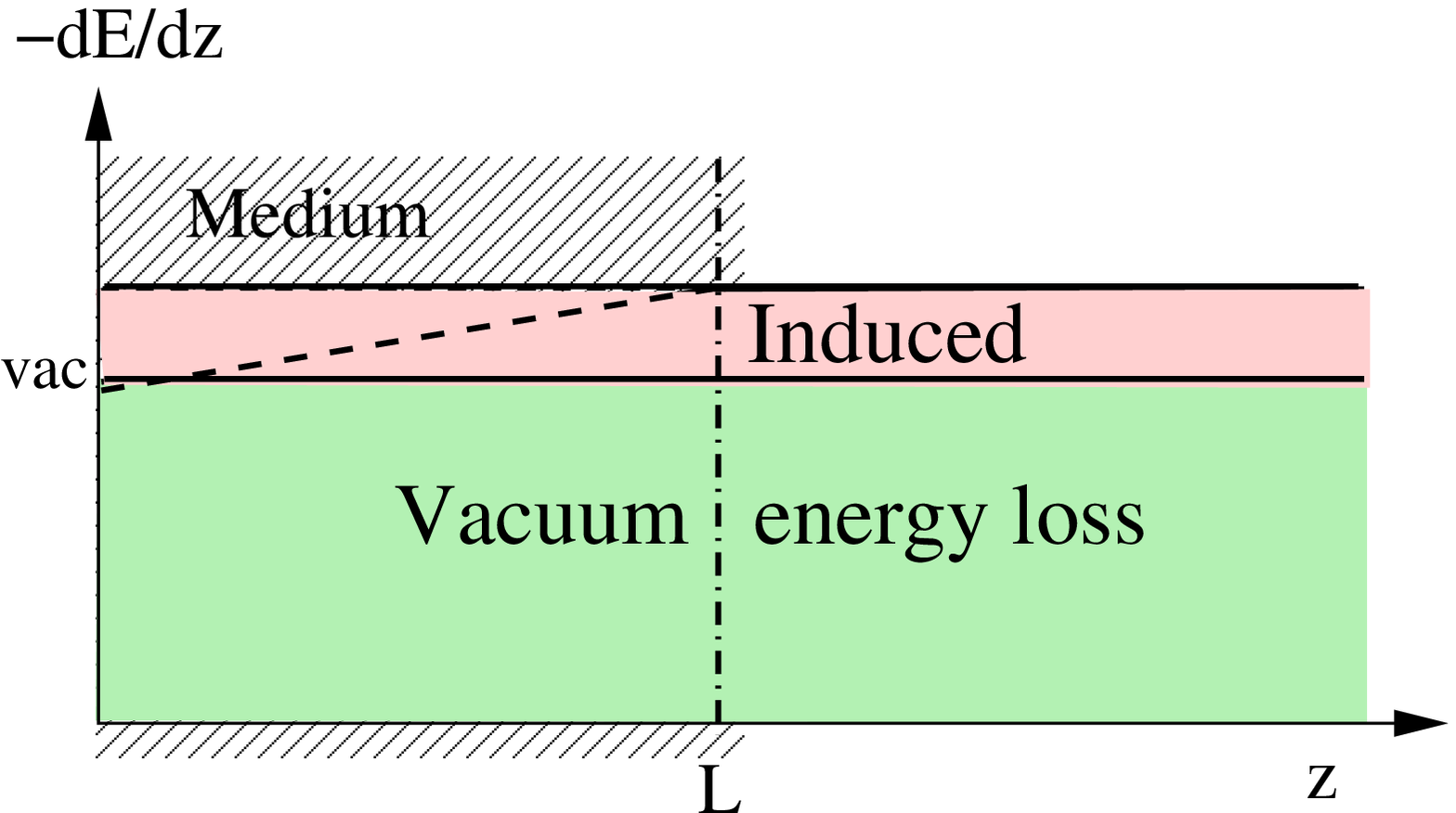}
 \caption{Exaggerated induced energy loss started at the maximal rate
which otherwise is reached only at the medium back surface.}
 \label{eloss2}
 \end{minipage}
 \end{figure}

Let us make the production length (\ref{30}) much longer than the nuclear
size, for instance by increasing the energy. Thus, gluon radiation
continues through the nucleus and outside. The rate of induced energy
loss linearly rises with the pathlength up to the medium surface
as is shown in Fig.~\ref{eloss1}.
 {\it What happens afterwards?}
According to the Landau-Pomeranchuk principle, radiation at longer
times does not resolve the structure of the interaction at the
initial state. What is important is the accumulated kick, and it
dies not matter whether it was a single or multiple kicks.
Therefore, the vacuum energy loss is continuing with a constant rate
increased due to final state interactions.

Lacking a good knowledge of the hadronization dynamics, one can impose an
upper bound for the medium-induced suppression. This bound can be
calculated precisely with no ad hoc procedures.

Let us increase the amount of induced energy loss assuming that its rate
does not rise up to the maximal value near the medium surface, but starts
with this maximal rate from the very beginning, as is illustrated in
Fig.~\ref{eloss2}. Since the induced energy loss is increased, the
resulting suppression of leading hadrons can only be enhanced.

We arrive at a constant rate of energy loss corresponding to hadronization
in vacuum, but with increased scale $Q^2\Rightarrow Q^2+\Delta p_T^2$. The
scale dependence of the fragmentation function can be calculated
perturbatively via of DGLAP equations
 \beq
\widetilde D^h_i(z_h,Q^2) = D^h_i(z_h,Q^2)
+ \frac{\Delta p_T^2}{Q^2}
\sum\limits_j\int\limits_{z_h}^1 \frac{dx}{x}
P_{ji}[x,\alpha_s(Q^2)] D^h_j(z_h/x,Q^2).
\label{70}
 \eeq
 The medium induces a harder scale which makes the energy loss more
intensive.  The difference is the induced energy loss which is $\propto
\Delta p_T^2$ and present implicitly in the DGLAP.\\[-7mm]
 \begin{figure}[htb]
\begin{minipage}[t]{65mm}
 \includegraphics[width=60mm]{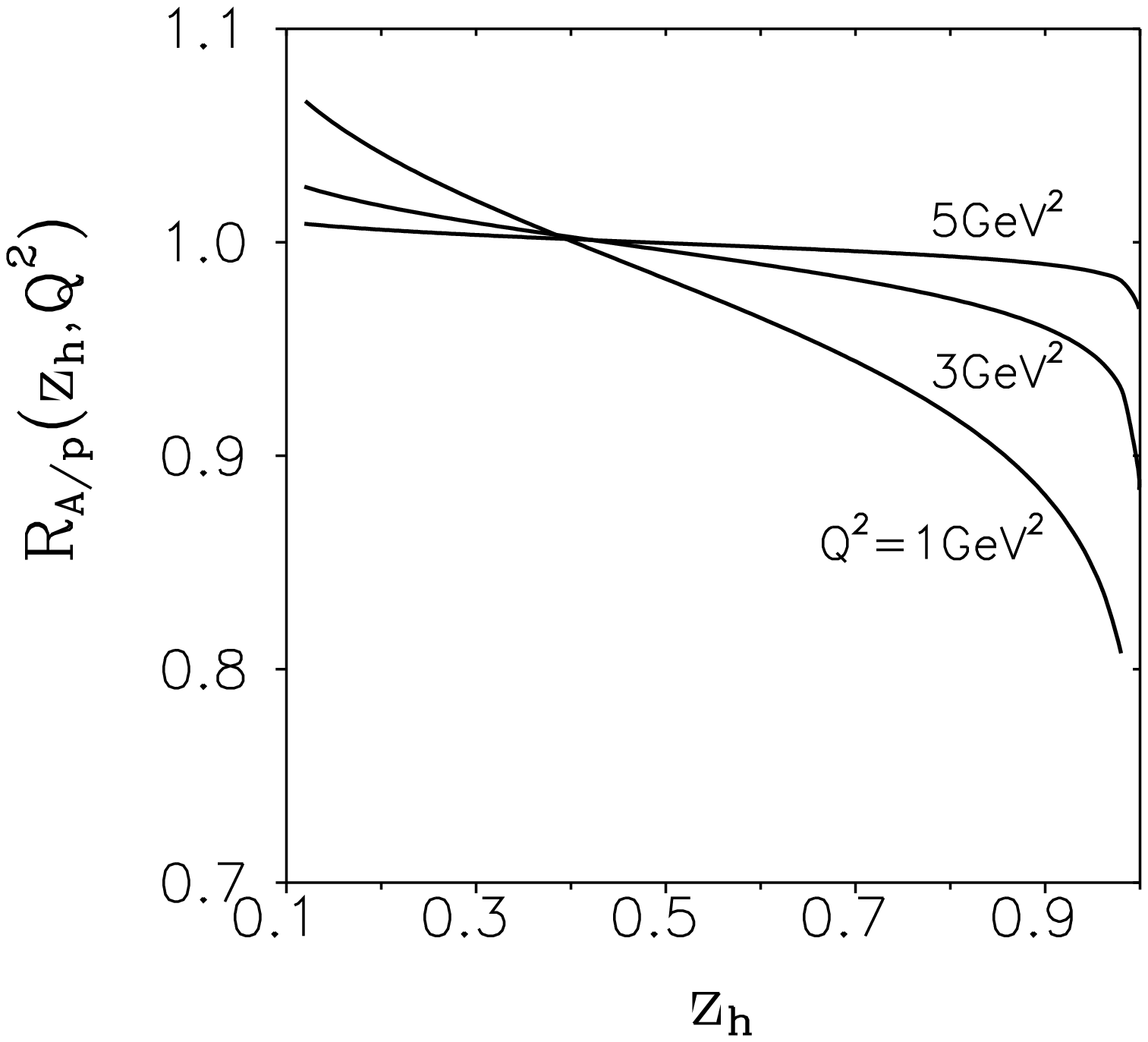}
\caption{Ratio of the nuclear-modified to vacuum fragmentation
functions calculated for {\it lead}. The modification is far too small in
comparison with data.}
\label{ratio}
\end{minipage}
\hspace{\fill}
\begin{minipage}[t]{75mm}
 \includegraphics[width=70mm]{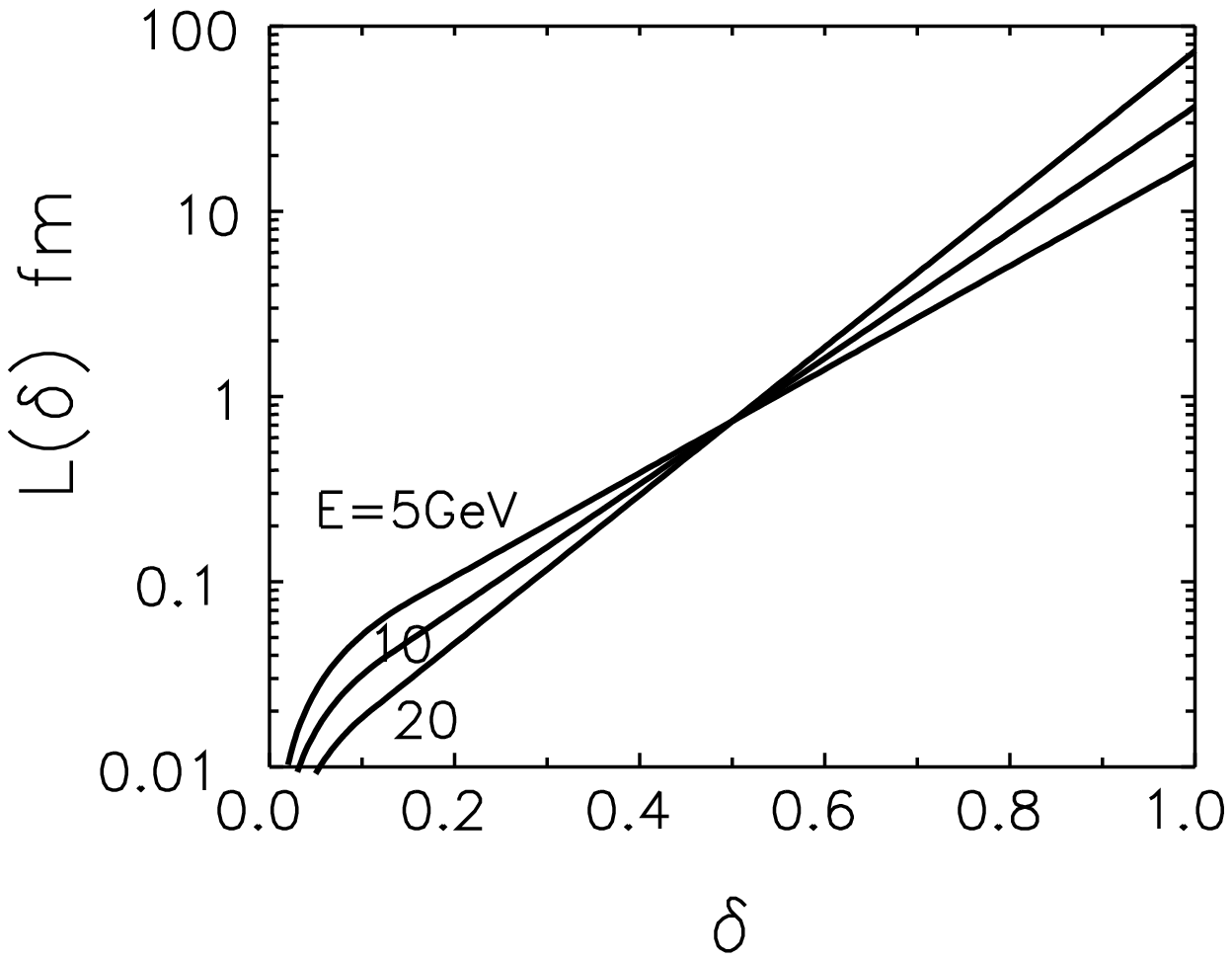}
 \caption{The path length needed to radiate fraction $\delta$ of
the total vacuum energy loss. Three curves correspond to different jet
energies, $E=5,\ 10$ and $20\GeV$.}
 \label{delta}
\end{minipage}
 \end{figure}

\section{Time evolution of a high-\boldmath$p_T$ jet}

{\it How much energy is radiated over path length $L$?}
 \beq
\Delta E(L) =
E\int\limits_{\Lambda^2}^{Q^2}
dk^2\int\limits_0^1 dx\,x\,
\frac{dn}{dx\,dk^2}\
\Theta\left( L-l_c\right),
\label{80}
 \eeq
 where
 \beq
l_c=\frac{2Ex(1-x)}{k^2}\ ;\ \ \ \ \ \
\frac{dn}{dx\,d^2k}
=\frac{\gamma}{x\,k^2}\nonumber\ ;\ \ \ \ \ \
\gamma=\frac{3\alpha_s}{\pi^2}
\label{90}
 \eeq

 The rate of energy loss is constant for each interval of $k^2$,
 \beq
\frac{dE}{dL\,dk^2} =
{1\over2}\,\gamma
\label{100}
 \eeq
 Radiation of gluons with given transverse momentum $k$ is continuing with
the constant rate $\gamma/2$ until the maximal length
$L_{max}(k^2)=2E/k^2$ is reached.

The total energy radiated over this maximal path length is
 \beq
\Delta E_{tot}=\int\limits_{\Lambda^2}^{Q^2}
dk^2\,{1\over2}\,\gamma\,\frac{2E}{k^2}=
\gamma\,E\,\ln\frac{Q^2}{\Lambda^2}
\label{110}
 \eeq

{\it How long does it take to radiate fraction $\delta$ of the total
emitted energy?} The answer depends on how large is $\delta$. For $Q=E$
 \beqn
L&=&\delta\,\frac{4}{E}\,
\ln\frac{E}{\Lambda}
\ \ \ \ \ \ \ \ \ \ \ \ \ \ \
{\rm if}\ \  \delta<1/\ln\left(\frac{Q^2}{\Lambda^2}\right)
\nonumber\\
L&=&\frac{2}{Ee}\,
\left(\frac{E}{\Lambda}\right)^{2\delta}
\ \ \ \ \ \ \ \ \ \ \ \ \
{\rm if}\ \  \delta>1/\ln\left(\frac{Q^2}{\Lambda^2}\right)
\label{120}
 \eeqn

Thus, {\it more than a half of the total energy is lost within $1\fm\,$!}
So, the color neutralization, or production length shrinks
with the jet energy.

Notice that this result does not contradict the fact that the
mean time of radiation of a gluon is long and rises with jet energy,
 \beq
\la l_c\ra =
\int\limits_{\Lambda^2}^{Q^2}
dk^2\int\limits_0^1 dx\,
\frac{dn}{dx\,dk^2}\,l_c(x,k^2)
= \frac{E}{\Lambda^2}\
\frac{1}{\ln(Q/\Lambda)\,
\ln(Q\Lambda/4E^2)}
\label{130}
 \eeq

The medium suppression factor $R_{AA}(p_T)$ is a result of the
interplay of two phenomena which act in opposite directions: as
$l_p$ shrinks with $p_T$, the amount of induced energy loss reduces,
and this should lead to a rising $R_{AA}(p_T)$.

However, contraction of the production length makes the path
available for absorption of the colorless pre-hadron longer. This
leads to a reduction of $R_{AA}(p_T)$. Usually attenuation caused by
absorption is quite a strong effect, however one should incorporate
it with precaution.

\section{Summary}
\begin{itemize}

\item
 Production of leading hadrons in hard reactions involves two stages
of time development: (i) propagation of a parton though the medium
accompanied with vacuum and induced gluon radiation; (ii)
perturbative color neutralization followed by evolution and attenuation of
the (pre)hadron in the medium.

\item
 Theoretical tools describing both stages are well developed and do not
need ad hoc fits to the data to be explained.

\item
 The production length of leading hadrons is controlled by coherence of
radiated gluons and energy conservation

\item
 $p_T$-broadening is a sensitive probe for the production length

\item
 Shortness of the production (color neutralization) length is the main
source of nuclear suppression of leading hadrons observed in DIS. There is
no room for induced energy loss.

\item
 Maximizing the induced energy loss one can reach a calculable upper bound
for the modification of the fragmentation function. It shows that the
effects of induced energy loss are far too weak to explain the
observed nuclear suppression of leading hadrons.

\item
 The time scale of vacuum gluon radiation in high-$p_T$ jets is very
short, less than $1\fm$. The production time of leading pre-hadrons
is even shorter by a factor of $(1-z_h)$.

\end{itemize}

\medskip

{\bf Acknowledgments:} We are grateful to Hans-J\"urgen Pirner for
numerous inspiring and informative discussions. This work was
supported in part by Fondecyt (Chile) grants, numbers 1030355, 1050519,
1050589, by DFG (Germany)  grant PI182/3-1, and by the Slovak Funding
Agency, Grant No. 2/4063/24.


\begin{thebibliography}{9}

\bibitem{star} STAR collaboration, J.~Adams et al., Phys. Rev. Lett.
{\bf 91} (2003) 172302.

\bibitem{phenix} PHENIX collaboration, S.S.~Adler et al., Phys. Rev.C{\bf
69} (2004) 034910.

\bibitem{kn} B.Z.~Kopeliovich and F.~Niedermayer, Sov. J. Nucl. Phys.
{\bf 42} (1985) 504; Yad. Fiz. {\bf 42} (1985) 797.

\bibitem{knph} B.Z.~Kopeliovich, J.~Nemchik, E.~Predazzi, A.~Hayashigaki,
Nucl. Phys. A{\bf 740} (2004) 211.

\bibitem{feri} F.~Niedermayer, {\it Phys.Rev.} {\bf D34} (1986) 3494.

\bibitem{knp} B.Z.~Kopeliovich, J.~Nemchik and E.~Predazzi, Proceedings
of the workshop on Future Physics at HERA, ed. by G.~Ingelman,
A.~De~Roeck and R.~Klanner, DESY 1995/1996, v. 2, 1038 (nucl-th/9607036);
Proceedings of the ELFE Summer School on Confinement physics, ed. by
S.D.~Bass and P.A.M.~Guichon, Cambridge 1995, Editions Frontieres, p. 391
(hep-ph/9511214).

\bibitem{bms} S.J.Brodsky, Ch.T.~Munger and Ivan~Schmidt,
Phys. Rev. D{\bf 49} (1994) 3228.

\bibitem{berger} E.~Berger, Z. Phys. {\bf C4} (1980) 289.

\bibitem{ihep} V.D.~Apokin et al., Sov. J. Nucl. Phys. 46 (1987) 877;
Yad. Fiz. 46 (1987) 1482.

\bibitem{kz} B.Z.~Kopeliovich and B.G.~Zakharov, Yad. Fiz. {\bf 46} (1987)
1535; Phys. Lett. {\bf B264} (1991) 434

\bibitem{hans} P.~Chiappetta, H.J.~Pirner, Nucl. Phys. {\bf B291} (1987)
765.

\bibitem{dhk} J.~Dolejsi, J.~H\"ufner and B.Z.~Kopeliovich,
Phys. Lett. B312 (1993) 235.

\bibitem{jkt} M.B.~Johnson, B.Z.~Kopeliovich and A.V.~Tarasov,
Phys. Rev. {\bf C63} (1991) 035203.

\bibitem{knst} B.Z.~Kopeliovich, J.~Nemchik, A.~Sch\"afer and
A.V.~Tarasov, Phys. Rev. Lett. 88 (2002) 232303.

\bibitem{zkl} B.Z.~Kopeliovich, L.I.~Lapidus, and A.B.~Zamolodchikov,
Sov. Phys. JETP Lett. 33 (1981) 595; Pisma v Zh. Exper. Teor. Fiz.
33 (1981) 612.

\bibitem{joel} P.L.~McGaughey, J.M.~Moss, and J.-Ch.~Peng, NUCOLEX 99,
Wako, Japan, 1999 (hep-ph/9905447); Ann. Rev. Nucl. Part. Sci. {\bf 49},
217 (1999).

\bibitem{pt-new} M.B.~Johnson et al., hep-ph/0606126.

\bibitem{will} CLAS Collaboration, K.~Hafidi et al., Proposal to PAC30,
JLAB, 2006.

\bibitem{knps} B.Z.~Kopeliovich, J.~Nemchik, H.J.~Pirner, and Ivan~Schmidt,
paper in preparation.

\bibitem{acardi} A.~Accardi, D.~Grunewald, V.~Muccifora, and H.J.~Pirner,
Nucl. Phys. {\bf A761} (2005) 67.

\bibitem{hermes2} HERMES Collaboration: A. Airapetian et al., Eur. Phys. J.
{\bf C20} (2001) 479 (hep-ex/0012049); A.~Airapetian et al.,
hep-ex/0307023, to appear in Phys. Lett. {\bf B}.

\end{thebibliography}
\end{document}